\documentclass[pre,showpacs,twocolumn,longbibliography]{revtex4-1}

\usepackage{lipsum}
\usepackage{hyperref}

\usepackage{color}
\usepackage[usenames,dvipsnames]{xcolor}
\usepackage{amsmath,amsthm,amssymb}
\usepackage{graphicx}
\usepackage{epsfig}
\usepackage{dcolumn}
\usepackage{bm}
\usepackage{mathrsfs}
\usepackage{multirow}
\usepackage[all]{xy}
\usepackage{pbox}
\usepackage{verbatim}

\usepackage{xr}
\newcommand{\er}[1]{Eq.~\eqref{#1}}
\newcommand{\ers}[2]{Eqs.~(\ref{#1}-\ref{#2})}

\newcommand{\Ers}[2]{Equations~(\ref{#1}-\ref{#2})}

\newcommand{\Eraa}[3]{Equations~(\ref{#1}), (\ref{#2}) and (\ref{#3})}

\def\(({\left(}
\def\)){\right)}
\def\[[{\left[}
\def\]]{\right]}

\newcommand{\be}{\begin{equation}}
\newcommand{\ee}{\end{equation}}
\newcommand{\ben}{\begin{eqnarray}}
\newcommand{\een}{\end{eqnarray}}
\newcommand{\beq}{\begin{equation}}
\newcommand{\eeq}{\end{equation}}

\newcommand{\la}{\langle}
\newcommand{\ra}{\rangle}

\newcommand{\e}{{\text{e}}}

\begin{document}

\title{Fluctuating hydrodynamics, current fluctuations and hyperuniformity in boundary-driven open quantum chains}

\author{Federico Carollo, Juan P. Garrahan, Igor Lesanovsky and Carlos P\'erez-Espigares}
\affiliation{School of Physics and Astronomy}
\affiliation{Centre for the Mathematics and Theoretical Physics of Quantum Non-Equilibrium Systems,
University of Nottingham, Nottingham, NG7 2RD, UK}

\date{\today}

\begin{abstract}
We consider a class of either fermionic or bosonic non-interacting open quantum chains driven by dissipative interactions at the boundaries and study the interplay of coherent transport and dissipative processes, such as bulk dephasing and diffusion.  Starting from the microscopic formulation, we show that the dynamics on large scales can be described in terms of fluctuating hydrodynamics (FH).  This is an important simplification as it allows to apply the methods of macroscopic fluctuation theory (MFT) to compute the large deviation (LD) statistics of time-integrated currents. In particular, this permits us to show that fermionic open chains display a third-order dynamical phase transition in LD functions.  We show that this transition is manifested in a singular change in the structure of trajectories: while typical trajectories are diffusive, rare trajectories associated with atypical currents are ballistic and hyperuniform in their spatial structure.
We confirm these results by numerically simulating ensembles of rare trajectories via the cloning method, and by exact numerical diagonalization of the microscopic quantum generator.  
\end{abstract}

\maketitle 

\section{Introduction}
There is much interest nowadays in understanding the collective macroscopic behaviour of non-equilibrium quantum systems that emerges from their underlying microscopic dynamics.  This includes problems of thermalisation \cite{Polkovnikov2011,Gogolin2016,Dalessio2016,Essler2016,Vasseur2016} and of novel non-ergodic \cite{Nandkishore2015} and driven phases \cite{Moessner2017} in quantum many-body systems, issues which also have started to be addressed experimentally \cite{Langen2015,Schreiber2015,Choi2016,Zhang2016,Bordia2016b,Choi2016b}.  Given that in practice interaction with an environment, while sometimes controllable, is always present, an important question is to what extent the interplay between coherent dynamics and dissipation influences the collective properties of such quantum non-equilibrium systems \cite{Jezouin2013,Nandkishore2014,Levi2016,Fischer2016,Medvedyeva2016,Luschen2016,Monthus2017}. 

The hallmark of a driven non-equilibrium system is the presence of currents.  Recently there has been important progress in the description of current bearing quantum systems with two complementary approaches.  One corresponds to the study of quantum quenches where two halves of a system are prepared initially in different macroscopic states, and where the successive non-equilibrium evolution displays stationary bulk currents associated with transport of conserved charges \cite{Bernard2015,Bernard2016,Bernard2016b,Castro-Alvaredo2016,Bertini2016,Ljubotina2017}.  Another corresponds to studies of one-dimensional spin chains coupled to dissipative reservoirs at the boundaries and described by quantum master equations \cite{Prosen2011XXZ,Znidaric2011b,Znidaric2010,Znidaric2010b,Znidaric2011,Eisler2011,Temme2012,Buca2014,Znidaric2014,Znidaric2014b,ilievski2014exact,prosen2015matrix,ilievski2016dissipation,Karevski2016,XXMonthus2017,Popkov2017}.  While the transport depends on the precise nature of the processes present in the system, the studies above find that in appropriate long wavelength and long time limits the dynamics can be described in terms of effective hydrodynamics.  A central question is  to what extent the non-equilibrium dynamics of systems where there is an interplay between quantum coherent transport and dissipation displays behaviour similar to that of classical driven systems \cite{Derrida2007}. 

Here we address the statistics of currents in
driven dissipative quantum systems and the spatial structure that is associated with rare current fluctuations.  We consider quantum chains for either fermionic or bosonic 
particles driven dissipatively through their boundaries.  We also allow for dephasing and/or dissipative hopping in the bulk.  Previous studies of similar systems have shown that with bulk dephasing
and/or dissipative hopping transport is diffusive, while in the absence of both it is ballistic \cite{Znidaric2010b,Znidaric2011,Eisler2011,Temme2012}.  
We show that the large scale dynamics of these systems in general admits a hydrodynamic description in terms of macroscopic fluctuation theory (MFT) \cite{Bertini2015}. In particular, we show that  quantum stochastic trajectories can be described at the macroscopic level in terms of fluctuating hydrodynamics (FH) \cite{Spohn1991}. This macroscopic dynamics corresponds to that of the classical symmetric simple exclusion (SSEP) for fermionic chains, and to the classical symmetric inclusion processes (SIP) for bosonic chains \cite{Giardina2007,Giardina2010,Carinci2013,Baek2016JSM}.

The effective classical MFT/FH description in turn allows us to obtain the large deviation (LD) \cite{Touchette2009} statistics of time-integrated currents \cite{Derrida2007,Hurtado2014,Lazarescurev2015}.  The cumulant generating function, or LD function, plays the role of a dynamical free-energy, and its analytic structure reveals the phase behaviour of the dynamics \cite{Bodineau2005,Garrahan2007,Lecomte2007,Garrahan2010,Hurtado2011,Gambassi2012,Espigares2013,Manzano2014,Tsobgni2016,Tizon2016,LecomteDPT2016,Lazarescu2017}.  We find that in fermionic open chains the LD function displays a third-order  transition between 
phases with very different transport dynamics.  Similar LD criticality was found for classical SSEPs with open boundaries \cite{Imparato2009,Lecomte2010}.  We show that this transition is manifested in a singular change in the structure of the steady state, from  diffusive for dynamics with typical currents, to ballistic and {\em hyperuniform} \cite{Torquato2016} (i.e., local density is anticorrelated and large scale density fluctuations are strongly suppressed) for dynamics with atypical currents.  

\section{Models and effective hydrodynamics}
\label{secquant} 
We consider a class of one-dimensional bosonic or fermionic non-interacting quantum systems of $L$ sites, weakly coupled to an environment, and connected 
at its first and last sites to density reservoirs. In the Markovian regime, we assume the evolution of a system operator $X$ to obey the following Lindblad equation \cite{Lindblad1976,Gorini1976}, 
\begin{equation}
\partial_tX_t = i[H,X_t]+{\cal D}^*[X_t] ,
\label{L}
\end{equation}
where $H$ is a quadratic Hamiltonian
\begin{equation}
H= \sum_{h=1}^{M}J_h\sum_{k=1}^{L-h}\left(a^\dagger_{k+h}a_k+a^\dagger_{k}a_{k+h}\right) ,
\label{H}
\end{equation}
with $a^\dagger_k,a_k$ denoting (bosonic or fermionic) creation and annihilation operators acting on site $k$.
Here $J_h$ is the coherent hopping rate between a given site $k$ and the sites $k\pm h$. The largest hopping distance $M$ is assumed to be finite and independent of $L$,
with $M=1$ corresponding to only nearest-neighboring hops.
The superoperator ${\cal D}(\cdot) := \sum_{\mu} V_{\mu} (\cdot) V_{\mu}^{\dagger} - \frac{1}{2} \{ V_{\mu}^{\dagger} V_{\mu} , (\cdot) \}$ is a Lindblad dissipator modelling the coupling to the environment; $V_{\mu}$ are {\em jump operators}, and $\{\cdot,\cdot\}$ stands for the anticommutator.  The dissipator has three contributions,
${\cal D} := {\cal D}_1 + {\cal D}_L + {\cal D}_{\rm bulk}$.  The first two corresponds to boundary driving terms where particles are introduced at site $1$ and $L$ with rates $\gamma_{1,L}^{\rm in}$ and removed with rates $\gamma_{1,L}^{\rm out}$, cf.~\cite{Znidaric2014,Znidaric2014b},
\begin{align}
{\cal D}^*_{1,L}[X]& := \gamma_{1,L}^{\rm in}
\left(
a_{1,L} X a_{1,L}^\dagger-\frac{1}{2}\{a_{1,L} a_{1,L}^\dagger, X\}
\right)
+
\nonumber
\\
&+
\gamma_{1,L}^{\rm out}
\left(
a_{1,L}^\dagger X a_{1,L}-\frac{1}{2}\{a_{1,L}^\dagger a_{1,L}, X\}
\right)
\, .
\label{D1L}
\end{align} 
The third contribution accounts for dissipation in the bulk, 
including site dissipation with rate $\gamma$ and dissipative hopping with rate $\varphi$, cf.~\cite{Temme2012,Eisler2011}, 
\begin{align}
{\cal D}^*_{\rm bulk}[X] := &
\gamma \sum_{k=1}^L\left(n_k X n_k-\frac{1}{2}\{n^2_k,X\}\right)+ 
\label{Dbulk}
\\
&
+\frac{\varphi}{2}\sum_{k=1}^{L-1}\left([[L^\dagger_k,X],L_k]+[[L_k,X],L_k^\dagger]\right)\, ,
\nonumber
\end{align}
with $n_k=a^\dagger_k a_k$ being the $k$-th site particle number operator, and $L_k=a^\dagger_{k+1}a_k$.  

To derive the effective macroscopic description we consider the average occupation on each site
$\la n_m \ra_t$, which from \ers{L}{Dbulk} obeys a {\em continuity equation},
\begin{equation}
\partial_t \la n_m \ra_t=- \la\sum_{h=1}^M \left(j_{h,m}^{\rm co}-j^{\rm co}_{h,m-h}\right)+\left(j_m^{\rm dis}-j^{\rm dis}_{m-1}\right)\ra_t\, ,
\label{cont}
\end{equation}
where $j^{\rm co}_{h,m} := -i \, J_h(a^\dagger_{m+h}a_m-a^\dagger_{m}a_{m+h})$, and $j^{\rm dis}_m := \varphi(n_{m}-n_{m+1})$, are the different current contributions: $j_{h,m}^{\rm co}$ is the coherent, and thus quantum in origin, particle current between sites $m$ and $m+h$, while $j_m^{\rm dis}$ is the analogous of the {\em stochastic} current in SSEPs. 
The next step is to rescale space and time by suitable powers of the chain length $L$ to get meaningful equations in the thermodynamic limit. The correct rescaling is the diffusive one, in which the macroscopic space and time variables are given by $x := m / L \in[0,1]$, and $\tau := t / L^2 $. In the new time-coordinate, the evolution of the average occupation number is implemented by
\begin{equation}
\partial_\tau \la n_m \ra_\tau=-L^2 \la \sum_{h=1}^M\left(j_{h,m}^{\rm co}-j^{\rm co}_{h,m-h}\right)+\left(j_m^{\rm dis}-j^{\rm dis}_{m-1}\right)\ra_\tau\, .
\label{cont2}
\end{equation}
In order to derive an equation for the average density one needs to focus on the quantum current $j^{\rm co}_{h,m}$, and to understand what is its contribution in the diffusive scaling. Neglecting terms which are not contributing in the rescaled space-time framework, (see Appendix \ref{AppA} for a detailed discussion on the hydrodynamic limit), one has the following time-derivative for the current 
\begin{equation}
\partial_\tau \la j_{h,m}^{\rm co}\ra_\tau \approx L^2\Big[2J_h^2\left\la n_m-n_{m+h}\right\ra_\tau-\tilde{\gamma}\la j_{h,m}^{\rm co}\ra_\tau\Big]\, ,
\label{Xcurr}
\end{equation}
where $\tilde{\gamma}=\gamma+2\varphi>0$ is the total dephasing rate. Formally integrating the above equation, neglecting exponentially decaying terms, one has 
$$
\la j_{h,m}^{\rm co}\ra_\tau \approx 2 J_h^2 L^2 \int_0^\tau du \, \e^{-\tilde{\gamma}L^2(\tau-u)}\left\la n_m-n_{m+h}\right\ra_u;
$$
for very large $L$, $L^2 \e^{-\tilde{\gamma}L^2(\tau-u)}$ converges, under integration, to a Dirac delta, (see Eq. \eqref{deltalike} in Appendix \ref{AppA}), and thus the quantum contributions to the current become $\la j_{h,m}^{\rm co}\ra_\tau\approx \frac{2J_h^2}{\gamma+2\varphi}\left\la n_{m}-n_{m+h}\right\ra_\tau$.  Substituting this into \eqref{cont2}, and recasting the various contributions, one finds 
\begin{equation}
\begin{split}
\partial_\tau \la n_m \ra_\tau&\approx\varphi  L^2 \la n_{m+1}-2n_m+n_{m-1}\ra_\tau +\\
&+\frac{2L^2 }{\tilde{\gamma}}\sum_{h=1}^M J_h^2\, \la n_{m+h}-2n_m+n_{m-h}\ra_\tau\, .
\end{split}
\end{equation}
The above expectations are proportional to finite-difference second-order derivatives. This means that when introducing the macroscopic density $\rho_{\tau}(x) := \la n_{m = x L}\ra_{t = \tau L^2}$, defined on the rescaled macroscopic space, in the large $L$ limit, one obtains the hydrodynamic equation 
\begin{equation}
\partial_\tau \rho_\tau(x)= D \, \partial_x^2 \rho_\tau(x)\, ,
\label{diff}
\end{equation}
where the effective diffusion rate reads
\begin{equation}
D := \left(\varphi+\frac{2}{\gamma+2\varphi}\sum_{h=1}^MJ_h^2\, h^2\right)\, .
\label{diff2}
\end{equation}
Dissipation at the chain ends, \er{D1L}, set the boundary conditions on the density $\rho_\tau(0) = {\varrho}_0$ and $\rho_\tau(1) = {\varrho}_1$ for all $\tau$ (see Appendix \ref{AppA}),
\begin{equation}
{\varrho}_0 := \frac{\gamma_1^{\rm in}}{\gamma_1^{\rm out}\pm\gamma_1^{\rm in}}, \quad {\varrho}_1 := \frac{\gamma_L^{\rm in}}{\gamma_L^{\rm out}\pm\gamma_L^{\rm in}}\, ,
\label{bc}
\end{equation}
where plus is for fermionic systems, while the minus, restricted to the case $\gamma_{1,L}^{\rm out}-\gamma_{1,L}^{\rm in}>0$, is for bosonic ones. Such restriction is necessary in the bosonic case in order to achieve a convergence of the boundary conditions.  \Eraa{diff}{diff2}{bc} encode the diffusive hydrodynamics of both fermionic and bosonic chains governed by \er{L}. This result agrees with previously studied special cases: for example, in the tight-binding ($M=1$) fermionic case without
dissipative hopping, Eq. (7) reduces to the diffusive rate of  \cite{Znidaric2014b}, while
for dissipative hopping without dephasing to the one found in \cite{Temme2012}.
%
%
The macroscopic dynamics \ers{diff}{bc}, thus unifies previous findings and predicts the behaviour for more general Hamiltonians, including also the bosonic case.
For non-quadratic Hamiltonians ({\it interacting systems}), our procedure cannot be directly applied as it is not straightforward to close the differential equations for the densities in the hydrodynamic limit. However, for particular limiting cases, 
it is possible to make predictions on how microscopic interactions modify the hydrodynamic behavior of the system through a diffusion rate which in general depends on the density profile. In the next section we discuss one of these cases.

\section{Example of Hamiltonian with interactions}
We consider a fermionic system whose bulk dynamics is implemented by 
$$
\partial_t X=i[H_{\rm{Int}},X]+\gamma\sum_{k=1}^L \left(n_kXn_k-\frac{1}{2}\{n_k,X\}\right),
$$
where the Hamiltonian, $H_{\rm{Int}}=H+H_V$ has also an interaction term
\begin{equation*}
H=J\sum_{k=1}^{L-1}\left(a^\dagger_{k+1}a_k+a^\dagger_ka_{k+1}\right),\quad H_V=V\sum_{k=1}^{L-1}n_k\, n_{k+1}\, .
\end{equation*}
The driving at the chain's ends is treated in the same way as in the previous section, providing the same density at the boundaries in terms of the injection and removal rates \eqref{bc}. We are interested in deriving a hydrodynamic equation for the above system in the limit of strong dephasing and strong interaction, i.e.~$|V|,\gamma\gg |J|$, as well as in shedding light on the role played by interactions in the coarse-grained description. To this aim it proves convenient to introduce the projector $\mathcal{P}$, whose action consists in projecting operators onto the subspace of diagonal elements in the number basis. Namely, given $X$ a monomial in creation and annihilation operators, one has $\mathcal{P}X=X$, if $X$ is diagonal in the number basis (e.g.~$X=a^{\dagger}_ma_m$), and $\mathcal{P}X=0$ otherwise (e.g.~$X=a^{\dagger}_ka_m$ with $k\ne m$).



\begin{figure*}
   \includegraphics[scale=0.5]{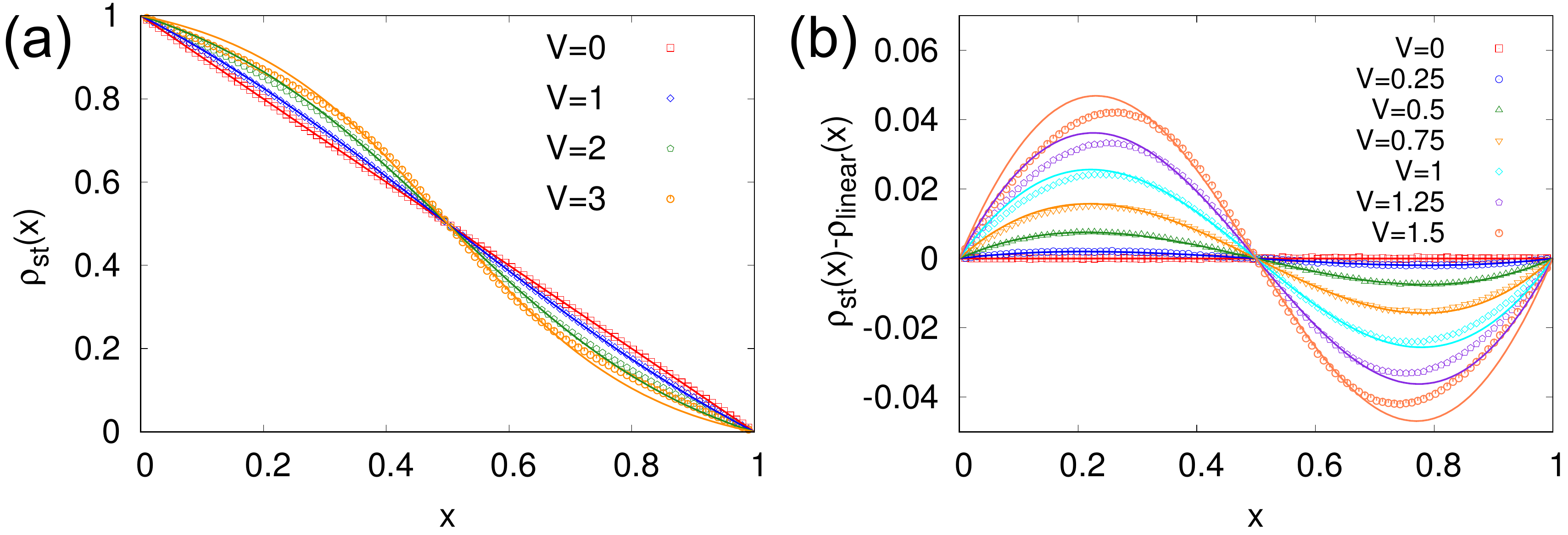}
  \caption{\small (Color online) Stationary density profiles for $\rho_0=1$, $\rho_1=0$, $J=1$ and $\gamma=1$ for increasing values of the interaction strength $V$. Solid lines correspond to the analytical density profile derived from equations \eqref{eqProfV} and \eqref{eqDProfV} while points correspond to the numerical results obtained by simulating the microscopic dynamics given by equation \eqref{DISINT} for $L=100$. (a): Stationary density profiles. (b): Difference between the stationary density profiles and the linear one $\rho_{\rm{linear}}(x)=\rho_0+(\rho_1-\rho_0)x$. 
}\label{fig1SM}
\end{figure*}

In the strong dephasing and interaction approximation, one obtains the following effective dynamical map \cite{vznidarivc2015relaxation,everest2017role,Monthus2017,XXMonthus2017} 
$$
\partial_t X=-\mathcal{P}\mathcal{H}\circ \frac{1}{\mathcal{D}_0}\circ\mathcal{H}\mathcal{P}X\, ,
$$
where $\mathcal{H}[X]=i[H,X]$ and $\mathcal{D}_0[X]=i[H_V,X]+\gamma\sum_{k=1}^L \left(n_kXn_k-\frac{1}{2}\{n_k,X\}\right)$. Considering the time-derivative of a generic $n_m$ in the bulk, one gets
\begin{equation}
\partial_t n_m=\Gamma_m\left(n_{m+1}-n_m\right)-\Gamma_{m-1}\left(n_m-n_{m-1}\right)\, ,
\label{DISINT}
\end{equation}
where
$$
\Gamma_{m}=\frac{2J^2\gamma}{\gamma^2+V^2(n_{m+2}-n_{m-1})^2}\, .
$$
This dynamics resembles the one of stochastic exclusion processes, although jump rates are non-trivial and feature an explicit dependence on the particle configuration \cite{everest2017role}. In the mean-field approximation, consisting in neglecting correlations $\langle n_h n_k\rangle\approx \langle n_h\rangle\langle n_k\rangle$, and considering the macroscopic density profile $\rho_\tau(x)$, with $x=\frac{m}{L}$, and the rescaled time $\tau=L^{-2} t$, one obtains 
\begin{equation*}
\begin{split}
\partial_\tau\rho_\tau(x)\approx L^2\Bigg[&\Gamma(x)\left(\rho_\tau(x+\frac{1}{L})-\rho_\tau(x)\right)+\\
&-\Gamma(x-\frac{1}{L})\left(\rho_\tau(x)-\rho_\tau(x-\frac{1}{L})\right)\Bigg]\, ,
\end{split}
\end{equation*}
with $\Gamma(x)=\frac{2J^2\gamma}{\gamma^2+V^2F(x)}$, and 
\begin{equation*}
\begin{split}
F(x)=\rho_\tau(x+\frac{2}{L})+\rho_\tau(x-\frac{1}{L})-2 \rho_\tau(x+\frac{2}{L}) \rho_\tau(x-\frac{1}{L}) .
\end{split}
\end{equation*}
In the large $L$ limit the differential equation becomes 
\be
\partial_\tau \rho_\tau(x)=\partial_x\Big(D\big(\rho_\tau(x)\big)\partial_x \rho_\tau(x)\Big)\, ,
\label{eqProfV}
\ee
where the diffusion rate, $D(\rho(x)):=\lim_{L\to \infty} \Gamma(x)$, is given by
\be
D(\rho)=\frac{2J^2\gamma}{\gamma^2+2V^2(\rho-\rho^2)}\, .
\label{eqDProfV}
\ee
This shows that the microscopic interactions are reflected at the coarse-grained hydrodynamic level in a non-trivial dependence of the diffusive transport coefficient on the density field. In order to check the range of validity of the above hydrodynamic equation, we compare the analytical result for the stationary state $\rho_{st}(x)$ of Eq. \eqref{eqProfV} with numerical results obtained by performing continuous-time Monte Carlo simulations of Eq. \eqref{DISINT} --see Fig.~\ref{fig1SM}. As $\rho_{st}(x)$ depends only on the ratio $V/\gamma$, we have set $J=\gamma=1$ and studied its dependence on $V$. In Fig.~\ref{fig1SM}(a) the analytical profiles are displayed along with the numerical ones. For the sake of clarity, in Fig.~\ref{fig1SM}(b) deviations from the linear profile ($V=0$) are displayed; for small values of $V$ (up to $V=0.75$) a good agreement is achieved between theory and simulations. Increasing the value of $V$, we can observe how the numerical results start to deviate from the analytical predictions due to the failure of the mean-field approximation. 


%
\section{Stochastic quantum trajectories and fluctuating hydrodynamics}
We now come back to the original non-interacting model described by equation \eqref{L}, which is the system we consider in our study. The obtained diffusive character of the driven quantum chain opens up the possibility of applying MFT \cite{Bertini2015} for calculating the LD statistics of currents. This would represent a huge simplification as it reduces the non-trivial task of computing the current LD function to a variational problem.  \Ers{diff}{bc} provide an effective description at the macroscopic level of the exact microscopic master equation, \ers{L}{Dbulk}.  In order to derive a MFT description we need an equivalent macroscopic characterization of the corresponding fluctuating quantum trajectories.  In the ``input-output'' formalism the dynamics of a system operator $X_t$ in terms of both the system and the environment is described in terms of a quantum stochastic differential equation (QSDE) \cite{Gardiner2004} ,\begin{align}
d X = & 
i [H,X] dt + {\cal D}^*(X) dt 
\label{sde}
\\
&
+ \sum_{\mu}
\left( [ V^{\dagger}_{\mu} , X ] dB_{\mu}
+ dB_{\mu}^{\dagger} [ X, V_{\mu}]
\right) 
\, ,
\nonumber
\end{align}
where $\{ V_{\mu} \}$ are the jump operators in the Lindblad master equation, cf.~\ers{L}{Dbulk}, and 
$dB_{\mu},dB_{\mu}^{\dagger}$ are operators on the environment representing a quantum Wiener process and obeying the quantum Ito rules, $(dB_{\mu})^2 = (dB_{\mu}^{\dagger})^2 = dB_{\mu}^{\dagger} \, dB_{\mu}=0$ and $dB_{\mu} \, dB_{\mu}^{\dagger} = dt$. To understand how the quantum trajectories are described at the macroscopic level we consider the simpler case $\varphi=0$. In this case, jump operators in the bulk are diagonal in the number operator basis, so that the presence of the environment does not directly affect the evolution of the density. Indeed, the variation in time of a bulk number operator is given by 
\begin{equation}
dn_m=-\sum_{h=1}^M(j_{h,m}^{\rm co}-j_{h,m-h}^{\rm co})dt\, .
\end{equation}
The presence of the Wiener process is instead explicit in the stochastic evolution of the quantum current contributions; in this case one has 
\begin{equation*}
\begin{split}
dj_{h,m}^{\rm co}&\approx \left[2J^2_h\left(n_m-n_{m+h}\right)-{\gamma} j_{h,m}^{\rm co}\right]dt+\\
&+\sqrt{\gamma}\sum_{k=1}^L\left([n_k,j_{h,m}^{\rm co}]dB_k(t)+dB_k^\dagger(t)[j_{h,m}^{\rm co},n_m]\right)\, ,
\end{split}
\end{equation*}
where we are neglecting those terms that were not contributing to the hydrodynamic equation \eqref{diff} and that can be shown to be irrelevant also in this stochastic regime. The first step is to introduce the time rescaling; it is important to notice that this affects in different ways the two increments: while $dt=L^2 d\tau$, one has --as for all Wiener processes-- $dB(t)=LdB(\tau)$.
Thus, the rescaled time stochastic equation reads
\begin{equation*}
\begin{split}
dj_{h,m}^{\rm co}&\approx L^2 \left[2J^2_h\left(n_m-n_{m+h}\right)-{\gamma} j_{h,m}^{\rm co}\right]d\tau+\\
&+L\sqrt{\gamma}\sum_{k=1}^L\left([n_k,j_{h,m}^{\rm co}]dB_k(\tau)+dB_k^\dagger(\tau)[j_{h,m}^{\rm co},n_m]\right).
\end{split}
\end{equation*}
The first term in the above equation is nothing but the deterministic part already present in \eqref{Xcurr}, leading to equation \eqref{diff}. The remaining contribution instead, modifies the deterministic equation for the evolution of the macroscopic density, introducing an extra noisy term (see Appendix \ref{AppFH}). In particular, rescaling space and considering the large $L$ limit, one has that the stochastic macroscopic field $\hat{\rho}_\tau$ obeys the following Langevin equation
\begin{equation}
\partial_{\tau} \hat{\rho}_{\tau}(x) =  -\partial_x \hat{j}_{\tau}(x)
\, , 
\label{FH}
\end{equation}
where $\hat{j}_{\tau}(x) := -D\partial_x \hat{\rho}_{\tau}(x) + \xi_{\tau}(x)$ indicates the fluctuating current field. The coarse-grained macroscopic effects due to the presence of the quantum Wiener process are encoded in the Gaussian noise $\xi$, determining deviations from the average behavior. This zero-mean Gaussian noise is characterized by a covariance which, under a local equilibrium assumption for the global quantum state \cite{Spohn1991}, is given by (see Appendix \ref{AppFH})
\begin{equation}
\la \xi_{\tau}(x)\xi_{\tau'}(x') \ra = L^{-1}\sigma({\hat \rho}_{\tau}(x)) \, \delta(x-x') \, \delta(\tau-\tau') 
\, , 
\label{xi}
\end{equation}
where the {\em mobility} $\sigma(\rho)$ is a function of the density profile 
\begin{equation}
\sigma(\rho) = 2 D \, \rho (1 \mp \rho) \,\,\, \text{fermions/bosons}\,. 
\label{sigma}
\end{equation}
It important to stress the fact that the stochastic macroscopic fields $\hat{\rho}_\tau,\hat{j}_\tau$ represent a coarse-grained hydrodynamic description of the quantum trajectories given by equation \eqref{sde}. The structure of equation \eqref{FH} remains unchanged for $\varphi\neq0$; one only needs to consider the appropriate diffusive parameter $D$. 

Equal densities at the boundaries, ${\varrho}_0={\varrho}_1=\rho$, corresponds to equilibrium conditions, for which the stationary quantum state is a product thermal one. Thus one can easily compute the {\em compressibility}, $
\chi:= L^{-1}\sum_{h,\ell=1}^{L}\left(\la \,n_k\,n_{\ell}\ra -\la n_h \ra \la n_{\ell} \ra\right)$, which can be expressed in terms of the average occupation $\rho$ as  $\chi(\rho) = \rho (1 \mp \rho)$ (for fermions/bosons).  This means that the {\em Einstein relation} \cite{Spohn1991} connecting the linear response of the density to a perturbation - the mobility - to its spontaneous fluctuations in equilibrium - the compressibility - is obeyed, $\sigma(\rho) = 2 D \chi(\rho)$.  Notice that we have derived $\sigma(\rho)$ starting from the quantum trajectories; this extends to generic Hamiltonians the mobility found, by means of perturbation theory, for the tight-binding case with dephasing \cite{Znidaric2014b}.
Remarkably, the form of the mobility given by \er{sigma} shows that the fluctuating hydrodynamic behavior of these quantum systems is equivalent to the SSEP for fermions and to the SIP for bosons \cite{Giardina2007,Giardina2010,Carinci2013,Baek2016JSM}.


\noindent
\section{Current fluctuations, ballistic dynamics and hyperuniformity in fermionic chains} 
The fluctuating hydrodynamics of the quantum chain \eqref{FH} encodes the evolution of any possible realization $\{\hat{\rho},\hat{j}\}$ of the system. Therefore, not only stationary properties can be derived, but also the dynamical behavior associated with fluctuations and atypical trajectories. In the following, we focus on fermionic chains and study the statistics of
the empirical (i.e., time-averaged) total current, $\hat{q} := T^{-1}\int_0^1 dx \int_0^{T} d\tau \, \hat{j}_{\tau}(x)$, up to a macroscopic time $T=t/L^2$.  For long times we expect its probability to have a LD form, $P_t({q}) := \langle \delta(q - \hat{q}) \rangle \approx e^{- t \phi({q})}$, where $\phi(q)$ is the LD {\em rate function}.  The same information is encoded in the moment generating function, $Z_t(s) := \la e^{- s\,t q} \ra$, where $s$ is the {\em counting field} conjugate to $q$. $Z_t(s)$ can be interpreted as a dynamical partition function and allows one to define for each $s$ a new ensemble of trajectories with biased probability, the so-called $s$-ensemble \cite{Garrahan2007}. Averages in this biased ensemble take the form $\la \cdot \ra_s=Z_t(s)^{-1}\la (\cdot) e^{-s\,t{\hat q}} \ra$, with $s=0$ being the original non-biased expectation.
The dynamical partition function has a LD form, $Z_t(s) \approx e^{t \theta(s)}$, where $\theta(s)$ is the {\em scaled cumulant generating function} (SCGF)
, and is related to $\phi(q)$ via a Legendre transform \cite{Touchette2009}. 
The SCGF plays the role of a dynamical free-energy, whose non-analytic behavior accounts for dynamical phase transitions. These correspond to singular changes in the trajectories sustaining atypical values of different observables. As shown in  \cite{Imparato2009}, 
the SSEP undergoes a third-order dynamical phase transition for current fluctuations at $s=0$. This is reflected in the following limit of the current SCGF obtained in Ref.~\cite{Imparato2009}, featuring a discontinuity in the third derivative:
\begin{align}
\tilde{\theta}(s):= & \lim_{L \to \infty} \frac{\theta(s)}{L} 
= & \frac{\sigma s^2}{2}+
\frac{D\sqrt{2}}{24\pi}\left|\frac{\sigma'' \sigma}{D^2}\right|^{3/2}|s|^3\, ,
\label{tilpsi}
\end{align} 
where $\sigma=\sigma(\rho_{\rm opt})$ and $\sigma''=\sigma''(\rho_{\rm opt})$ with $\rho_{\rm opt}$ the time-independent {\em optimal} profile \cite{Shpielberg2016} sustaining the atypical current associated with $s$. While this dynamical phase transition was already predicted in \cite{Imparato2009}, its physical implications at the level of the trajectories is still lacking. In the following, we shall unveil the nature of this transition: while dynamics leading to typical empirical currents is diffusive, the one associated to atypical currents is ballistic and with hyperuniform spatial structure.  

Firstly, we analytically show this change of behavior by means of the FH approach; then we compare our theoretical predictions with extensive numerical simulations 
of the rare trajectories of the SSEP, finding good agreement for the largest system size we could reach ($L=64$).

Finally, as predicted by \eqref{FH}, we shall show how the FH results correctly describe the hydrodynamics of the quantum models introduced in section \ref{secquant} 
through the exact numerical computation of the LD properties of a quantum spin chain.



\subsection{Structure factor for rare trajectories}


\begin{figure*}
  \includegraphics[scale=0.5]{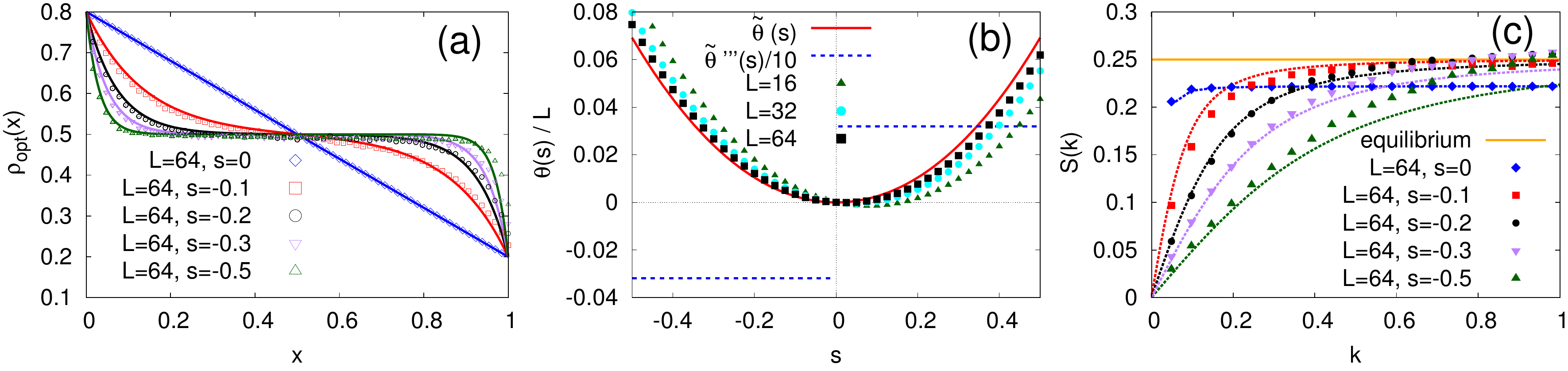}
  \caption{\small (Color online) {\bf Dynamical transition and hyperuniformity  in the one-dimensional SSEP with open boundaries.} Here ${\varrho}_0=0.8$, ${\varrho}_1$=0.2 and $D=1$. Lines correspond to analytical predictions while symbols to numerical results. (a) Optimal profiles for four different values of $\lambda=sL$ with $L=64$. The numerical results confirm that the profiles tend to $\rho = 1/2$ as $\lambda=sL$ increases. (b) SCGF ${\tilde \theta} (s)$ given by \er{tilpsi} (solid red line) together with its third derivative (dashed blue line). (c) Static structure factor $S(k,0)$ for different values of the bias $s$.}\label{fig1}
\end{figure*}

In order to determine the different dynamical phases signalled by \eqref{tilpsi}, one needs to study the dynamical structure factor, cf.~\cite{Jack2015,Karevski2017}. 
In general, this quantity, $S(k,t) := L^{-1} \, \la \delta \tilde{n}_k(0)\delta \tilde{n}_k^*(t) \ra$,
is defined in terms of the spatial Fourier transform, $\delta{\tilde n}_k(t)$,
of the microscopic particle fluctuations, $\delta n_m(t):=n_m(t)-\la n_m \ra_t$. 
By taking into account the open geometry of the system, this is given by,
$$
\delta \tilde{n}_k(t)=\sqrt{2}\sum_{h=1}^L \sin(k\,h)\delta n_h(t)\,
$$
where $k=\frac{\pi}{L}r,\, r=1,2,\dots L-1$. By substituting this in the definition of the structure factor one obtains
\begin{equation}
S(k,t)=\frac{2}{L}\sum_{\ell,h=1}^L\sin(k\,\ell)\sin(k\, h) C_{\ell h}(t)\, ,
\label{micstruc}
\end{equation}
with $C_{\ell h}(t)=\langle \delta n_{\ell}(0) \delta n_h(t)\rangle_s$ being the second cumulant of densities at site $h$ and $k$ averaged in the $s$-ensemble \cite{Garrahan2007}.
While at this microscopic level the computation of density-density correlations is just possible for small system sizes, one can still derive a closed expression for the structure factor by exploiting the macroscopic approach of fluctuating hydrodynamics. In the large $L$ limit,  
one can approximate summations with integrals, obtaining the following relation
$$
\delta\tilde{n}_k(t)=\sqrt{2}L\int_0^1dx\, \sin(p\, x)\delta\rho_\tau(x)=L\delta\tilde{\rho}_{\tau}(p)\, ,
$$
with $ p=L\,k$, and $\delta\tilde{\rho}_{\tau}(p)$ being the Fourier sine transform of $\delta\rho_\tau(x)$, which encodes the macroscopic density fluctuations around the optimal profile $\rho_{\rm opt}(x)$ for a given value of $s$. Hence we can write $S(k,t)$ in terms of macroscopic quantities, $S(k,t) = L \, {\cal S}(p,\tau)$ with $p = Lk$, where ${\cal S}(p,\tau) = \la \delta \tilde{\rho}_0(p) \delta \tilde{\rho}_\tau^*(p) \ra_s$ and $\delta\tilde{\rho}_\tau(p)$. These averages over the $s$-ensemble can be cast in a path-integral representation \cite{Janssen1992,Tailleur2008},  $\la \cdot \ra_s=\e^{-t \theta (s)}\int  D \rho D {\bar{\rho}} \, (\cdot) \e^{-L\int dx d\tau {\cal L}[\rho,{\bar{\rho}}]}$, where $\bar{\rho}$ is a response field, and the Lagrangian reads 
\begin{equation}
\mathcal{L}[\rho,\bar{\rho}]=i\bar{\rho}\left(\partial_{\tau} \rho-D\partial_x^2 \rho\right)-\lambda D\partial_x\rho-\frac{\sigma(\rho)}{2}\left(i\partial_x\bar{\rho}-\lambda\right)^2\, ;
\nonumber
\end{equation}
$\lambda=sL$ is a macroscopic counting field, associated with the average current per site \cite{Appert-Rolland2008,Imparato2009}.
 To evaluate ${\cal S}(p,\tau)$ we need the quadratic expansion of the Lagrangian in terms of $\delta\rho$ and $\delta\bar\rho$,  
\begin{equation}
\begin{split}
&\mathcal{L}_2[\delta \rho,\delta\bar{\rho}]=i\delta\bar{\rho}\left(\partial_{\tau}\delta\rho-D\partial_x^2\delta \rho\right)+
\frac{\sigma}{2}\left(\partial_x\delta\bar{\rho}\right)^2+\\
&-\frac{\sigma^{\prime\prime}}{4}\left(i\partial_x\bar{\rho}_{opt}-\lambda\right)^2(\delta \rho)^2-i\sigma^{\prime}(i\partial_x\bar{\rho}_{opt}-\lambda)\delta \rho\partial_x\delta\bar{\rho}\, .
\end{split}
\label{secordL}
\end{equation}
Since in general the coefficients in this quadratic expansion are space-dependent, the Gaussian integral is non-trivial.  However, in the equilibrium case, ${\varrho}_0={\varrho}_1=1/2$, these coefficients become constant, and the integration  needed for ${\cal S}(p,\tau)$ becomes straightforward in terms of Fourier modes (see Appendix \ref{AppSF}).  Remarkably, 
for large $\left|\lambda\right|$ (i.e. finite $s$), regardless of the density at the boundaries, the optimal profile tends to maximize  $\sigma(\rho)$, thus adopting the half-filling configuration $\rho(x) = {1}/{2}$, except for vanishingly small regions at the boundaries \cite{Karevski2017,Imparato2009}.  This allows for the computation of the structure factor associated with the rare trajectories, which reads  (see Appendix \ref{AppSF} for details),
\begin{equation}
S(k,t) 
= \sigma k^2
\frac{\exp\left(-\frac{t}{2} 
\sqrt{4 D^2 k^4-2 s^2\sigma^{\prime\prime}\sigma k^2}\right)
}
{\sqrt{4 D^2 k^4-2 s^2\sigma^{\prime\prime}\sigma k^2}}\, .
\label{dynSF}
\end{equation}
Notice that the static structure factor associated to equal time density-density correlations is obtained by taking $S(k,t=0)$. 

It follows from \eqref{dynSF} that for typical dynamics, $s=0$, at equilibrium with $\varrho_0=\varrho_1=\frac{1}{2}$, the structure factor is diffusive, $S(k,t) \propto \exp(- D k^2 t)$.  In contrast, and more interestingly, for $s \neq 0$ and for any value of the density at the boundaries, we get, 
\begin{equation}
S(k,t) \sim \frac{\sigma |k|}{\sqrt{-2s^2\sigma^{\prime\prime}\sigma}}
\exp\left(-\frac{|k| \, t }{2} \sqrt{-2 s^2\sigma^{\prime\prime}\sigma} \right) \, ,
\end{equation}   
for small $k$. This means two things: (i) Dynamics associated with empirical currents away from the typical value have ballistic dynamical scaling, $t \approx L^z$ with dynamical exponent $z=1$. (ii) For small $k$ the structure factor vanishes linearly in $|k|$, i.e.~large-scale density fluctuations are suppressed and the system becomes spatially {\em hyperuniform} \cite{Torquato2016}.  This shows that for driven fermions the most efficient way to generate dynamics with atypical values of the current is by means of a singular change to a hyperuniform spatial structure, similarly to what occurs in the SSEP with periodic boundaries \cite{Jack2015}. 

\subsection{Simulation and numerical results} 

The theoretical predictions are based on the assumption that the fluctuations around the optimal profile are small, so that the Lagrangian can be approximated with its quadratic expansion. To validate this assumption we have performed advanced numerical simulations of the classical stochastic model corresponding to Eq. \eqref{FH} with $\sigma(\rho)=2D\rho(1-\rho)$, namely, the boundary-driven SSEP. These are obtained via the cloning method in continuos time with 1000 clones \cite{Giardina2006,Lecomte2007b,Tailleur2009,Giardina2011}; this method efficiently generates rare trajectories by means of population dynamics techniques similar to those of quantum diffusion Monte Carlo. Figs.~\ref{fig1}(a)-(b) display the numerical density profiles and the associated numerical SCGF, showing good agreement with the MFT predictions. In Fig.~\ref{fig1}(c) we observe how the numerical static structure factor follows the theoretical prediction, especially for small values of $|s|$.  Nevertheless, simulations allow us to explore larger values of $s$, showing that a hyperuniform spatial structure persists. These computational results confirm the validity of the analytical predictions obtained via the FH approach.

\begin{figure}[t!]
\includegraphics[width=8cm]{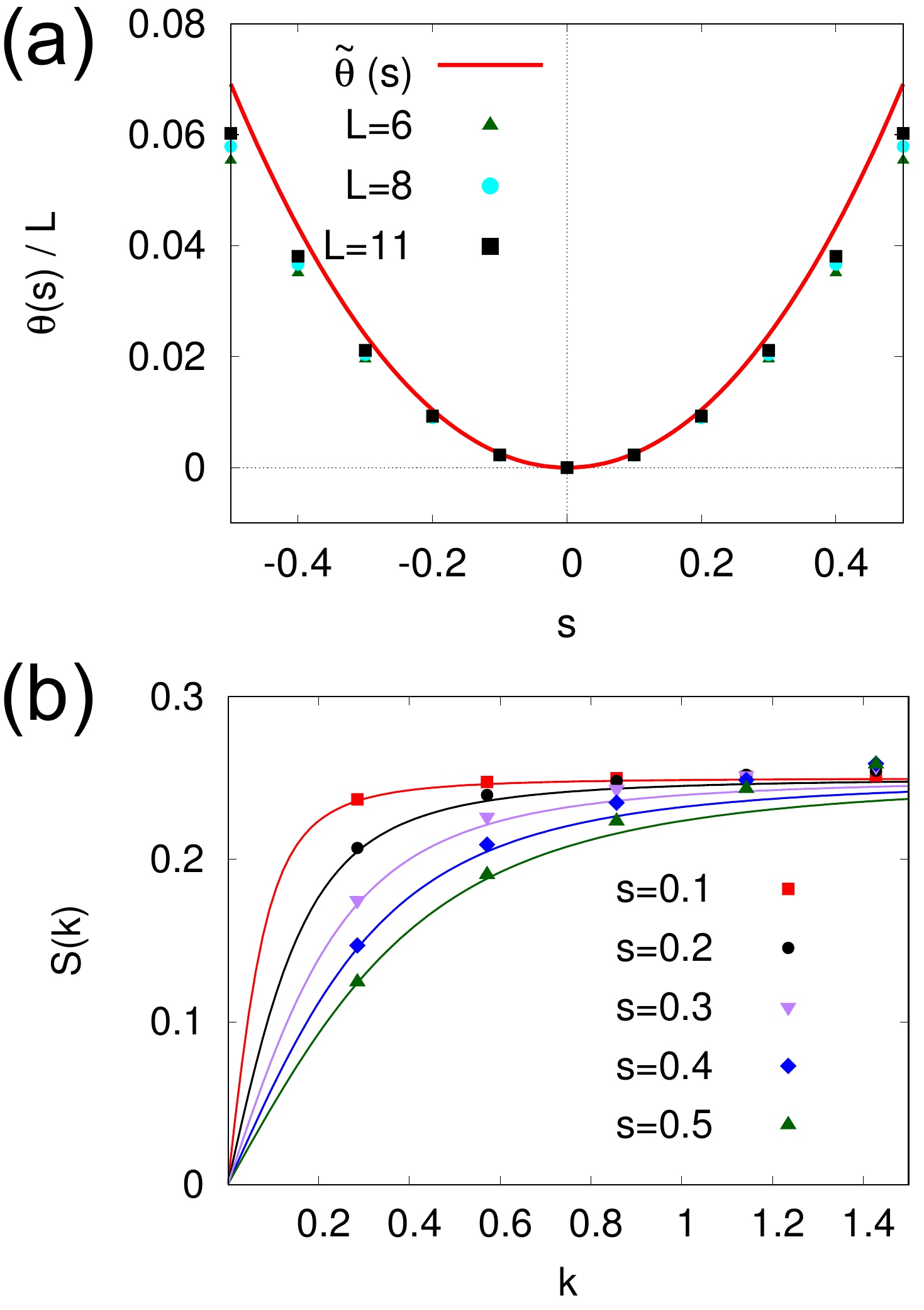}
\caption{\small (Color online) Top panel: SCGF for a fermionic quantum system given by the largest eigenvalue of Eq. \eqref{qtilt} for different system sizes (symbols), together with the SCGF, ${\tilde \theta} (s)$,  predicted by fluctuating hydrodynamics \er{tilpsi} (solid red line).  Bottom panel: Structure factor for $L=11$ 
for different values of $s$, along with the fluctuating hydrodynamic prediction (solid lines).}
\label{figQscgfSF}
\end{figure}

Since we have shown that the quantum trajectories of the models under consideration admit a macroscopic description in terms of Eq. \eqref{FH}, the previous analytical predictions hold as well for the boundary-driven fermionic quantum chains. In order to confirm the validity of our findings, we provide results from exact numerical diagonalization of the microscopic quantum tilted generator \cite{Garrahan2010}. We study the case of a fermionic quantum system given by Eq. \eqref{L}, with $M=J_1=\gamma_{1,L}^{\rm in/out}=1$, $\varphi=0$, and $\gamma=2$. Following \cite{Znidaric2014b,Znidaric2014}, the SCGF for the current in a boundary-driven fermionic chain can be obtained computing the eigenvalues with the largest real part of the tilted generator 
\begin{equation}
\begin{split}
\mathcal{W}_\lambda[X]=&-i[H,X]+\mathcal{D}[X]+\\
+&(\e^\lambda-1)a_L X a_L^\dagger+(\e^{-\lambda}-1)a_L^\dagger X a_L\, .
\end{split}
\label{qtilt}
\end{equation}
We shall consider $\lambda=sL$, since we are interested in the total extensive current. Moreover, from the left and  right eigenmatrices associated with the largest real eigenvalue of $\mathcal{W}_\lambda$ one can construct the stationary state for each value of $\lambda$ \cite{Garrahan2010}. This allows one to compute the density-density correlations necessary to determine the structure factor in the s-ensemble. In Fig.~\ref{figQscgfSF}(a) we show the results for the SCGF for $L=6,8,11$, together with the static structure factor (Fig.~\ref{figQscgfSF}(b)) for $L=11$. This is the largest size we could reached with exact numerical diagonalization of this quantum problem. In the top panel, we can observe a good convergence towards the predicted SCGF with the system size. Remarkably, in the bottom panel the numerical results of the structure factor show a good agreement with the FH predictions already for $L=11$. 
These results confirm the equivalence between the fluctuating hydrodynamics of fermionic chains and SSEPs, predicted by Eq. \eqref{FH}.


\section{Conclusions} 

In this work we have unveiled the general hydrodynamic behaviour of a broad class of boundary driven dissipative quantum systems. These include non-interacting systems, for which our results apply in all dissipation regimes, and interacting systems in the limit of strong dephasing where interactions give rise to a density dependent diffusion coefficient. Furthermore, starting from an unravelling of the open quantum dynamics in terms of stochastic quantum trajectories we have derived, for the non-interacting case, an effective fluctuating hydrodynamic that describes fluctuations in microscopic trajectories at a coarse-grained level. Interestingly, this effective description is equivalent to the fluctuating hydrodynamics of classical simple exclusion/inclusion processes for fermionic/bosonic systems. Exploiting this analogy, we have shown that fermionic chains undergo a dynamical phase transition at the level of fluctuations, from a phase corresponding to typical diffusive dynamics when trajectories are conditioned on having typical values of (time-integrated) currents, to a phase with ballistic dynamics and hyperuniform spatial structure when trajectories are conditioned on atypical values of currents.  Our theoretical predictions are confirmed both by extensive numerical simulations of rare trajectories in the open classical SSEPs - in particular corroborating the dynamical structure factors obtained from the FH approach - and via exact numerical diagonalization of the quantum tilted generator - indicating the validity of the effective FH description of the open quantum chains. It would be interesting to experimentally probe our predicted phase transitions by monitoring current fluctuations in boundary-driven cold atomic lattice systems, for example via a variant of the experiment reported in Ref.~\cite{landig2015}.

\begin{acknowledgments}
This work was supported by EPSRC Grant No. EP/M014266/1, H2020 FET Proactive project RySQ (Grant No. 640378), and ERC Grant Agreement No. 335266 (ESCQUMA). We are also grateful for access to the University of Nottingham High Performance Computing Facility.
\end{acknowledgments}

\appendix
\section{Effective macroscopic description}
\label{AppA}

We present the derivation of the effective diffusive equation, \er{diff}, governing the dynamics of the coarse-grained particle density profile of the quantum chain. First of all, we provide a relation that will be extensively used throughout the derivation: it can be checked that
\begin{equation}
\lim_{L\to\infty}L^2\int_0^tdu\, \e^{-L^2\gamma(t-u)}f_L(u)=\frac{1}{\gamma}f_{\infty}(t)\, ,
\label{deltalike}
\end{equation}
whenever $\left\{f_L(u)\right\}_L$ is a sequence of bounded functions $\forall u>0$, converging in $L$; namely, $L^2$ times the exponential converges weakly (under integration) to a Dirac delta $\delta(t-u)$. \\

Given the quantum master equation $\partial_tX_t=i[H,X_t]+\mathcal{D}^*[X_t]$, we are interested in deriving an effective dynamics for the average density of particles in the rescaled coordinates, $\tau= t/L^2$, $\frac{m}{L}\to x\in [0,1]$.  The time-scale $t/L^2$ can be simply obtained by multiplying the action of the generator by a factor $L^2$,
$$
\partial_{\tau}X_\tau= L^2\left( i[H,X_\tau]+\mathcal{D}^*[X_\tau]\right).
$$
Regarding the spatial dimension, coarse-graining consists in mapping the $L$-site chain onto a line $\Lambda=[0,1]$, in such a way that the $m$-th site of the chain corresponds to the point $\frac{m}{L}$ in $\Lambda$. In order to account for this geometric mapping, one has to notice that for large $L$, the spacing between sites in $\Lambda$, equal to $\frac{1}{L}$, becomes infinitesimal. We now introduce a continuous description; namely, we interpret $a^\dagger_{m}$ as the creator of a particle in the one-dimensional box centred in $\frac{m}{L}$ of width $\frac{1}{L}$. Mathematically, by means of the continuous fields $\left\{\alpha_x,\alpha_{x}^\dagger\right\}_{x\in\Lambda}$, one has
$$
a^\dagger_{m}=\sqrt{L}\int_{U_m}dx\, \alpha_x^\dagger
$$
where $U_m$ is the domain of the box across the point $\frac{m}{L}\in\Lambda$; the multiplying factor $\sqrt{L}$ is needed to guarantee the commutation (anti-commutation) relations of the discrete bosonic (fermionic) operators, starting from the continuous ones for $\alpha_x,\alpha_x^\dagger$. Moreover, the integral can be approximated for large $L$ by
\begin{equation}
a^\dagger_{m}\sim \frac{1}{\sqrt{L}}\alpha^\dagger_{\frac{m}{L}}\, .
\label{dis-to-cont}
\end{equation}
With this at hand, we have all the ingredients to perform the hydrodynamic limit. We shall first consider the simplest case of a tight-binding Hamiltonian and then extend the result to more general situations.

\subsection{Nearest-neighbour coherent hopping $M=1$}
In this section we focus on the tight-binding Hamiltonian $H=J\sum_{k=1}^{L-1}\left(a^\dagger_{k+1}a_k+a^\dagger_ka_{k+1}\right)$; its action on quadratic operators reads
\begin{equation}
i\left[H,a^\dagger_m a_n\right]=iJ \left[\Delta_L^2\left(a^\dagger_{m}\right)a_n-a^\dagger_m\Delta_L^2\left(a_{n}\right)\right]\, \, ,
\label{Hquad}
\end{equation}
with 
$$
\Delta_L^2(X_m)=X_{m+1}-2X_m+X_{m-1}\, .
$$
For a generic bulk site $m$, the rescaled time-derivative of the expectation of the number operator $n_m=a^\dagger_ma_m$ can be cast in the following form 
\begin{equation}
\partial_\tau \langle n_m\rangle_\tau=-L^2\left(\langle j_m^{\rm co}-j^{\rm co}_{m-1}\rangle_\tau+\langle j_m^{\rm dis}-j^{\rm dis}_{m-1}\rangle_\tau\right)\, ;
\label{d-n}
\end{equation}
the operator $j_m^{\rm co}$ has the meaning of a coherent current through the sites $m$, $m+1$ and is defined as 
\begin{equation}
j^{\rm co}_m:=-iJ\left(a^\dagger_{m+1}a_m-a^\dagger_{m}a_{m+1}\right)\, ,
\end{equation} 
while 
$$
j^{\rm dis}_m:=\varphi\left(n_{m}-n_{m+1}\right)\, .
$$
To close the equation for the number operators, one needs to work on the current $j^{\rm co}_m$. In particular, given the dissipator $\mathcal{D}^*$ and by using  \eqref{Hquad}, its $\tau$ time-derivative reads
\begin{equation}
\partial_\tau j_m^{\rm co}=L^2 \left[2J^2\left(n_m-n_{m+1}\right)+J^2 Q_m-\tilde{\gamma} j_m^{\rm co}\right]\, ,
\label{d-current}
\end{equation}
with 
\begin{equation}
Q_m=a^\dagger_{m+2}a_m+a^\dagger_ma_{m+2}-a^\dagger_{m+1}a_{m-1}-a^\dagger_{m-1}a_{m+1}\, ,
\label{Q_m}
\end{equation}
and $\tilde{\gamma}=\gamma+2\varphi>0$.
The term $Q_m$ will be shown not to contribute in the hydrodynamic limit, but still its derivative needs to be considered:
\begin{equation}
\partial_\tau Q_m=L^2\left(-\tilde{\gamma} Q_m+i[H,Q_m]\right)\, ;
\label{ddd}
\end{equation}
we will study later the action of the Hamiltonian on this operator. Now, by formally integrating \eqref{d-current} and \eqref{ddd} and by substituting the result for $Q_m$ into the time-evolution of the coherent current, we get
\begin{equation}
\begin{split}
\langle j_m^{\rm co}\rangle_\tau&\sim -2J^2 L^2\int_0^\tau du\, \e^{-L^2\tilde{\gamma} (\tau-u)}\Delta_L \langle n_m\rangle_u+\\
&+L^4\int_0^\tau du\, \int_0^udv\, \e^{-L^2\tilde{\gamma} (\tau-v)}i\langle[H,Q_m]\rangle_v\, ,
\label{qu-curr}
\end{split}
\end{equation}
with $\Delta_L n_m=n_{m+1}-n_m$ and where we have neglected exponentially decaying terms in $L$.
To proceed, we substitute the above result, for $j_m^{\rm co},j_{m-1}^{\rm co}$, in equation \eqref{d-n}. By doing this and rearranging terms we find  
\begin{equation}
\begin{split}
&\partial_\tau \langle n_m\rangle_\tau=2J^2L^4\hspace{-0.1cm}\int_0^\tau\hspace{-0.1cm}du\,\e^{-L^2\tilde{\gamma}(\tau-u)}\Delta_L^2\langle n_m\rangle_u+\\
&+\varphi L^2\Delta_L^2\langle n_m\rangle_\tau -iJ^2L^6\int_0^\tau\int_0^ududv\,\e^{-L^2\tilde{\gamma}(\tau-v)}\langle P_m\rangle_v\, ,
\end{split}
\label{density}
\end{equation}
with $P_{m}=[H,{Q}_{m}-{Q}_{m-1}]$.
Considering relation \eqref{dis-to-cont} and noticing that $P_m$ is quadratic in bosonic/fermionic operators, we introduce the operator $\tilde P_{\frac{m}{L}}$, which is the quadratic operator resulting from $P_m$, just by replacing the discrete field with the continuous ones. In this way one is able to write the following differential equation for the expectation of $\eta_x=\alpha^\dagger_x\alpha_x$ 
\begin{equation}
\begin{split}
&\partial_\tau \langle \eta_{\frac{m}{L}}\rangle_\tau=2J^2L^4\int_0^\tau du\,\e^{-L^2\tilde{\gamma}(\tau-u)}\Delta_L^2\langle \eta_{\frac{m}{L}}\rangle_u+\\
&+\varphi L^2\Delta_L^2\langle \eta_{\frac{m}{L}}\rangle_\tau -iJ^2L^6\int_0^\tau
\int_0^ududv\,\e^{-L^2\tilde{\gamma}(\tau-v)}\langle \tilde{P}_{\frac{m}{L}}\rangle_v\, .
\end{split}
\label{d-cont}
\end{equation}
The term $L^2\Delta_L^2 \langle\eta_{\frac{m}{L}}\rangle_\tau$ represents a finite difference second derivative of the density, which in the large $L$ limit with $\frac{m}{L}\to x$ becomes 
\begin{equation}
\lim_{L\to\infty}\varphi\, L^2\Delta^2_L\langle \eta_{\frac{m}{L}}\rangle_\tau=\varphi \, \partial_x^2 \langle\eta_x\rangle_\tau\, .
\end{equation}
Similarly, taking also into account \eqref{deltalike}, one has
$$
\lim_{L\to\infty}2J^2L^4\hspace{-0.1cm}\int_0^\tau\hspace{-0.1cm}du\,\e^{-L^2\tilde{\gamma}(\tau-u)}\Delta_L^2\langle \eta_{\frac{m}{L}}\rangle_u=\frac{2J^2}{\tilde{\gamma}}\partial_x^2\langle\eta_x\rangle_\tau\, .
$$

It remains to show that the last term of the r.h.s of equation \eqref{d-cont} does not contribute in the large $L$ limit. Firstly, one can check that the following quantity is bounded:
$$
\lim_{L\to\infty}J^2L^4\int_0^\tau du\int_0^udv\,\e^{-L^2\tilde{\gamma}(\tau-v)}=C<\infty\, .
$$
As a consequence the modulus of the last term in \eqref{d-cont} 
\begin{equation}
I=\lim_{L\to\infty}\left|J^2L^6\int_0^\tau\int_0^ududv\,\e^{-L^2\tilde{\gamma}(\tau-v)}\langle \tilde{P}_{\frac{m}{L}}\rangle_v\right|
\end{equation}
can be bounded by
\begin{equation}
I\le C\lim_{L\to\infty}\max_{\forall t>0}\left\{L^2\left|\langle\tilde{P}_{\frac{m}{L}}\rangle_t\right|\right\}\, .
\end{equation}
To understand the contribution of the operator $\tilde{P}_{\frac{m}{L}}$, one needs to go back to the operator ${Q}_m$. The latter is made of the product of operators spaced by two lattice sites. For example, by using equation \eqref{Hquad}, one has  
\begin{equation}
i[H,a^\dagger_{m+2}a_m]=iJ\left(\Delta_L^2(a^\dagger_{m+2})\,a_m-a^\dagger_{m+2}\Delta_L^2(a_m)\right)\, ;
\end{equation}
then, by multiplying by $L^2$ and considering the spatial scaling one has
\begin{equation}
\begin{split}
\lim_{L\to\infty}L^2\langle\Delta_L^2(\alpha^\dagger_{\frac{m+2}{L}})\alpha_{\frac{m}{L}}-&\alpha^\dagger_{\frac{m+2}{L}}\Delta_L^2(\alpha_{\frac{m}{L}})\rangle_t=\\
&=\left(\langle \partial_x^2 \alpha_x^\dagger \, \alpha_x\rangle_t -\langle \alpha_x^\dagger \partial_x^2\alpha_x\rangle_t\right)\, .
\end{split}
\label{tqs}
\end{equation}
Due to the spatial coarse-graining, all terms of $[H,Q_m]$ give the same hydrodynamic contribution equal to \eqref{tqs}. Since $Q_m$ is made of two terms with positive sign and another two with negative one, the net result for the hydrodynamic limit of $\la [H,Q_m]\ra_{\tau}$ is zero. This implies that in the large $L$ limit $L^2 \la {\tilde P}_{\frac{m}{L}}\ra_t\to0$.
Thus, by defining $\rho_\tau(x)=\langle \eta_x\rangle_\tau$, we have shown that \eqref{d-cont} reads
\begin{equation}
\partial_\tau \rho_\tau(x)=\left(\varphi+\frac{2J^2}{\gamma+2\varphi}\right)\partial_x^2 \rho_\tau(x)\, ,
\label{det-diff}
\end{equation}
which corresponds to \er{diff} for $M=1$.

Such a differential equation needs to be provided with two boundary conditions; these are given by the extremal sites of the chain.
For the expectation value of the number operator of the first site $a^\dagger_1a_1$, one has 
\begin{equation}
\begin{split}
\partial_\tau \langle a^\dagger_1a_1\rangle_\tau =L^2\Big[&\gamma_{1}^{in} -(\gamma_{1}^{out}\pm\gamma_1^{in})\langle a^\dagger_1a_1\rangle_\tau+\\
+&iJ\langle a^\dagger_2a_1-a^\dagger_1a_2\rangle_\tau+\varphi\, \langle a^\dagger_2a_2-a^\dagger_1 a_1\rangle_\tau\Big]\, ,
\end{split}
\end{equation}
where the plus is for fermionic systems while the minus for bosonic ones. In the latter case, $\gamma_{1}^{out}-\gamma_1^{in}>0$ is needed for the convergence of the expectations. Formally integrating the above equation we get, neglecting exponentially decaying terms,
\begin{equation}
\begin{split}
\langle a_1^\dagger a_1\rangle_\tau&-\frac{\gamma_1^{in}}{\gamma_1^{out}\pm\gamma_1^{in}}\approx L^2\int_0^\tau du\,\e^{-L^2(\gamma_1^{out}\pm\gamma_1^{in})(\tau-u)}\times\\
&\times \left(iJ\langle a^\dagger_2a_1-a^\dagger_1a_2\rangle_u+ \varphi\, \langle a^\dagger_2a_2-a^\dagger_1 a_1\rangle_u\right)\, .
\end{split}
\end{equation}
The right-hand side of the above relation, using \eqref{deltalike}, going to the continuous description, and using that 
\begin{equation*}
\lim_{L\to\infty}\langle \alpha^\dagger_{\frac{2}{L}}\alpha_{\frac{1}{L}}-\alpha^\dagger_{\frac{1}{L}}\alpha_{\frac{2}{L}}\rangle_{u}=0,\,
\lim_{L\to\infty}\langle \alpha^\dagger_{\frac{2}{L}}\alpha_{\frac{2}{L}}-\alpha^\dagger_{\frac{1}{L}}\alpha_{\frac{1}{L}}\rangle_{u}=0\, ,
\end{equation*}
can be shown to go to zero in the large $L$ limit. Thus,
one finds that the left boundary density is given by
\begin{equation}
\varrho_0=\lim_{L\to\infty}\langle a^\dagger_{1}a_{1}\rangle_\tau=\frac{\gamma_1^{in}}{\gamma_1^{out}\pm\gamma_1^{in}}\, .
\label{bound0SM}
\end{equation}
Similarly, at the right boundary one has 
\begin{equation}
\varrho_1=\lim_{L\to\infty}\langle a^\dagger_{L}a_{L}\rangle_\tau=\frac{\gamma_L^{in}}{\gamma_L^{out}\pm\gamma_L^{in}}\,,
\label{bound1SM}
\end{equation}
provided ${\gamma_L^{out}\pm\gamma_L^{in}}>0$.

\subsection{Short-range coherent particle hopping ($M$ finite)}
The starting point in this case is the time-rescaled differential equation \eqref{cont2}.  As before, one needs to work on the generic quantum current contribution $j_{h,m}^{\rm co}$; its time-derivative can be casted in the following form
\begin{equation}
\begin{split}
\partial_\tau j_{h,m}^{\rm co}=L^2\Big[&2J_h^2\left(n_m-n_{m+h}\right)+J_h^2Q_{m}^h+\\
&+J_h\sum_{\ell\neq h}J_\ell Q_m^{h,\ell} -\tilde{\gamma}j_{h,m}^{\rm co}\Big]\, .
\end{split}
\label{d-currSH}
\end{equation}
The term $Q_m^h$ is a generalization of $Q_m$ of equation \eqref{Q_m}, and reads
$$
Q_m^h=a^\dagger_{m+2h}a_m-a^\dagger_{m+h}a_{m-h}-a^\dagger_{m-h}a_{m+h}+a^\dagger_m a_{m+2h}\, , 
$$
while, by defining  $\Delta_{L,\ell}^2(X_m)=X_{m+\ell}-2X_m+X_{m-\ell}$, $Q_m^{h,\ell}$ can be written as 
\begin{equation}
\begin{split}
Q_m^{h,\ell}&=\Delta_{L,\ell}^2(a^\dagger_{m+h})a_m-a^\dagger_{m+h}\Delta_{L,\ell}^2(a_m)+\\&-\Delta^2_{L,\ell}(a^\dagger_m)a_{m+h}+a^\dagger_m\Delta^2_{L,\ell}(a_{m+h})\, .
\end{split}
\label{Q_m^hl}
\end{equation}
One can show that, in the hydrodynamic limit, neither $Q_m^h$ nor $Q_m^{h,\ell}$, contribute to the differential equation. Regarding $Q_m^h$, this can be shown by performing analogous manipulations to the ones involved in the discussion of the term $Q_m$ in the previous case. For $Q_m^{h,\ell}$, we show below that the contribution is vanishing in the limit of large $L$. 

Let us focus on the first summand of the right-hand side of the above equation, written in terms of the continuous creation and annihilation operators $\alpha_x^\dagger,\alpha_x$ (see \eqref{dis-to-cont}) and multiplied by the time-rescaling factor $L^2$. One has, with $x=m/L$
$$
L^2\Delta_{L,\ell}^2\left(\alpha_{x+\frac{h}{L}}^\dagger\right)\alpha_x=\ell^2\frac{\alpha^\dagger_{x+\frac{h+\ell}{L}}-2\alpha^\dagger_{x+\frac{h}{L}}+\alpha^\dagger_{x+\frac{h-\ell}{L}}}{\left(\frac{\ell}{L}\right)^2}\alpha_x\,.
$$ 
In the large $L$ limit, considering that $h/L\to 0$ and $\ell/L\to 0$, this term becomes 
$$
\lim_{L\to\infty}L^2\Delta_{L,\ell}^2\left(\alpha_{x+\frac{h}{L}}^\dagger\right)\alpha_x=\ell^2\partial_x^2 \alpha^\dagger_x\, \alpha_x.
$$
Moreover, also the third term on the right-hand side of equation \eqref{Q_m^hl} converges to the same second order derivative obtained above, and, since it appears in $Q_{m}^{h,\ell}$ with a minus sign, it cancels the contribution given by the first term $Q_m^{h,\ell}$. The same happens for the remaining two terms. Therefore, formally integrating $j_{h,m}^{\rm co}$ and substituting the result in the equation for the number operator \eqref{cont2}, the hydrodynamic contribution from $Q_m^{h,\ell}$ vanishes. We thus have that the differential equation for the evolution of the density $\langle \eta_x\rangle =\langle \alpha^\dagger_x\alpha_x\rangle $, which reads
\begin{equation}
\begin{split}
\partial_\tau\langle\eta_{\frac{m}{L}}\rangle_\tau&=L^2\varphi\Delta_{L}^2(\langle\eta_{\frac{m}{L}}\rangle)+\\
&+2\sum_{h=1}^MJ_h^2L^4\hspace{-0.1cm}\int_0^\tau\hspace{-0.1cm}du\,\e^{-L^2\tilde{\gamma}(\tau-u)}\Delta_{L,h}^2\langle \eta_{\frac{m}{L}}\rangle_u\, .
\end{split}
\label{difeqSH}
\end{equation}
Taking into account relation \eqref{deltalike} and given that, with $x=\frac{m}{L}$,
$$
\lim_{L\to\infty}L^2\Delta_{L,h}^2\langle n_m\rangle_\tau=h^2\partial^2_x \langle \eta_x\rangle_\tau\,
$$
one obtains in the hydrodynamic limit, with $\rho_\tau(x)=\langle\eta_{x}\rangle_\tau$,
$$
\partial_\tau\rho_\tau(x)=\left[\varphi+\frac{2}{\gamma+2\varphi}\sum_{h=1}^MJ_h^2h^2\right]\partial_x^2\rho_\tau(x)\, ,
$$
which corresponds to \er{diff} for finite $M$.

\section{fluctuating hydrodynamics}
\label{AppFH}
We start from the time-rescaled equations 
\be
dn_m=-L^2\sum_{h=1}^M(j_{h,m}^{\rm co}-j_{h,m-h}^{\rm co})d\tau ;
\label{dnm}
\ee
and 
\begin{equation*}
\begin{split}
dj_{h,m}^{\rm co}&\approx L^2 \left[2J^2_h\left(n_m-n_{m+h}\right)-{\gamma} j_{h,m}^{\rm co}\right]d\tau+\\
&+L\sqrt{\gamma}\sum_{k=1}^L\left([n_k,j_{h,m}^{\rm co}]dB_k(t)+dB_k^\dagger(t)[j_{h,m}^{\rm co},n_m]\right)\, ,
\end{split}
\end{equation*}
where, in the latter, we have neglected the term $Q_m^h$ and $Q_{m}^{h,\ell}$ of \eqref{d-currSH} that, as in the deterministic case of the previous section, can be shown not to contribute to the effective equation.
By manipulating the noise term, the above equation can be rewritten as
\begin{equation*}
\begin{split}
dj_{h,m}^{\rm co}\sim L^2 \left[2J_h^2\left(n_m-n_{m+h}\right)-{\gamma} j_{h,m}^{\rm co}\right]d\tau\\
+L\sqrt{\gamma}J_hX_{h,m}dN_{h,m}(\tau)\, ,
\end{split}
\end{equation*}
with $X_{h,m}=a^\dagger_{m+h}a_m+a^\dagger_ma_{m+h}$ and
\begin{equation*}
\begin{split}
dN_{h,m}(\tau)=&-i\left(dB_{m+h}(\tau)-dB_{m+h}^\dagger(\tau)\right)+\\
&+i\left(dB_{m}(\tau)-dB_{m}^\dagger(\tau)\right)\, .
\end{split}
\end{equation*}
Integrating the above differential equation for $j_{h,m}^{\rm co}$, and substituting it in the equation \eqref{dnm} one finds, using relation \eqref{deltalike},
\begin{equation}
\begin{split}
dn_m&=\frac{2}{\gamma}\sum_{h=1}^MJ_h^2L^2\Delta^2_{L,h}(n_m)d\tau+\\
&-\sum_{h=1}^M\frac{J_h}{\sqrt{\gamma}}\left(X_{h,m}dN_{h,m}(\tau)-X_{h,m-h}dN_{h,m-h}(\tau)\right)\, ,
\end{split}
\end{equation}
where we have used \eqref{deltalike} and neglected the exponentially decaying term in $L$. 
Moving to the continuous coordinate given by \eqref{dis-to-cont}, again with $\eta_x=\alpha_x^\dagger\alpha_x,\, x=\frac{m}{L}$, one gets, with $\Delta_{L,h}(O_x)=O_x-O_{x-\frac{h}{L}}$,
\begin{equation}
\begin{split}
\partial_\tau\eta_x=&\frac{2}{\gamma}\sum_{h=1}^MJ_h^2L^2\Delta^2_{L,h}(\eta_x) +\\
-&L\sum_{h=1}^M\Delta_{L,h}\left(\frac{J_h}{\sqrt{L\gamma}}\tilde{X}_{h,x}\frac{d\nu_{h,x}(\tau)}{d\tau}\right) \, ,
\end{split}
\label{stoc}
\end{equation}
with $\tilde{X}_{h,x}$ being the analogous operator of $X_{h,m}$ but written in terms of the continuous fields $\alpha_x,\alpha^\dagger_x$; $d\nu_{h,x}$ is also the analogous operator of $dN_{h,m}(\tau)$, but written in term of the coarse-grained environment's operators $d\beta_x(\tau),d\beta_x^\dagger(\tau)$. In particular, the analogous relation to \eqref{dis-to-cont} holds
$$
dB_m^\dagger(\tau)\sim\frac{1}{\sqrt{L}}d\beta^\dagger_x(\tau)\, ,
$$
which is responsible, for the extra factor $\frac{1}{\sqrt{L}}$ in the round brackets of equation \eqref{stoc}. We see in \eqref{stoc} the same deterministic diffusion term of \eqref{difeqSH} (for $\varphi=0$) minus the first derivative of a noise term, that we denote by $\xi_\tau(x)$, 
$$
\xi_\tau(x)=\sum_{h=1}^M\frac{J_h}{\sqrt{L\gamma}}\, h\tilde{X}_{h,x}\frac{d\nu_{h,x}(\tau)}{d\tau}\, .
$$
The factor $h$ multiplying each term of the sum is due to the fact that $\Delta_{L,h}$ converges, in the hydrodynamic limit, to $h$ times the first derivative with respect to $x$.
Thus, the evolution equation for the fluctuating density $\hat{\rho}_\tau(x)=\langle \eta_x\rangle_\tau $ is given by 
$$
\partial_\tau\hat{\rho}_\tau(x)=D\partial_x^2\hat{\rho}_\tau(x)-\partial_x\xi_\tau(x)\, ,
$$
where the noise term $\xi_\tau(x)$ has a covariance 
$$
\langle \xi_\tau(x)\xi_{\tau'}(y)\rangle=\frac{\sigma (\hat{\rho}_\tau(x))}{L}\delta(x-y)\delta(\tau-\tau')\, ,
$$
$\sigma(\rho)=2D\rho(1\pm \rho)$, where the plus stands for bosons and the minus for fermions.
This covariance can be derived by directly computing 
\begin{equation*}
\begin{split}
\langle \xi_\tau(x)\xi_{\tau'}(y)\rangle=&\sum_{h,\ell=1}^M\frac{J_h J_\ell\,h \ell}{L\gamma}\times\\
&\times \left\la\tilde{X}_{h,x}\tilde{X}_{\ell,y} \frac{d\nu_{h,x}(\tau)}{d\tau}\frac{d\nu_{\ell,y}(\tau')}{d\tau'}\right\ra\, .
\end{split}
\end{equation*}
To understand the contribution of the operator $\tilde{X}_{h,x}\tilde{X}_{\ell,y}$ with $x=\frac{m}{L},\, y=\frac{n}{L}$, we look at the discrete original one 
$$
X_{h,m}X_{\ell,n}=\left(a^\dagger_{m+h}a_m+a^\dagger_m a_{m+h}\right)\left(a^\dagger_{n+\ell}a_n+a^\dagger_na_{n+\ell}\right);
$$
expanding the product one gets
\begin{equation*}
\begin{split}
X_{h,m}X_{\ell,n}&=a^\dagger_{m+h}a_m a^\dagger_{n+\ell}a_n+a^\dagger_m a_{m+h}a^\dagger_{n+\ell}a_n +\\
&+a^\dagger_{m+h}a_ma^\dagger_na_{n+\ell}+a^\dagger_m a_{m+h}a^\dagger_na_{n+\ell}.
\end{split}
\end{equation*}
Due to the presence of dephasing, damping quantum coherences on fastest time-scales than those of $\tau$, we assume a local ``thermal" equilibrium state for the infinitesimal domain across the bonds $x=\frac{m}{L},\, y=\frac{n}{L}$ \cite{Spohn1991} . This local equilibrium assumption implies that we have to consider only those terms in $X_{h,m}X_{\ell,n}$ giving a non-zero expectation over a free thermal equilibrium state. This happens only when $m=n$ and $h=\ell$, where one has 
$$
\langle X_{h,m}^2 \rangle \sim \langle n_m(1\pm n_{m+h})+n_{m+h}(1\pm n_m)\rangle
$$
Hence, the contribution of the coarsed-grained operator ${\tilde X}_{h,x}^2$ reads
$$
\langle \tilde{X}_{h,x}^2\rangle =2\hat{\rho}_\tau(x)\left(1\pm\hat{\rho}_\tau(x)\right)\, ,
$$ 
where the plus stands for bosons and the minus for fermions, and with $\hat{\rho}_\tau(x)$ being the fluctuating particle density.
Taking into account also the contribution of $\left\la \frac{d\nu_{h,x}(\tau)}{d\tau}\frac{d\nu_{h,y}(\tau')}{d\tau'}\right\ra$, one has that the non-vanishing terms are
$
\left\la\tilde{X}_{h,x}\tilde{X}_{h,y}\ \frac{d\nu_{h,x}(\tau)}{d\tau}\frac{d\nu_{h,y}(\tau')}{d\tau'}\right\ra=4\delta(x-y)\delta(\tau-\tau')\hat{\rho}_\tau(x)\left(1\pm\hat{\rho}_\tau(x)\right)
$
and thus, the full covariance of the noise $\xi_\tau(x)$ is given by 
\begin{equation*}
\begin{split}
\langle \xi_\tau(x)\xi_{\tau'}(y)\rangle=4\sum_{h=1}^M\frac{J_h^2\,h^2}{L\gamma}\hat{\rho}_\tau(x)&\left(1\pm\hat{\rho}_\tau(x)\right)\times\\
&\times  \delta(x-y)\delta(\tau-\tau')\, .
\end{split}
\end{equation*}
This means that, starting from the quantum stochastic master equation describing the quantum trajectories of the microscopic evolution, the equation governing the fluctuating hydrodynamics in the coarse-grained macroscopic description reads
$$
\partial_\tau \hat{\rho}_\tau(x)=-\partial_x \hat{j}_\tau(x)\, ,~~~\hat{j}_\tau(x)=-D\partial_x\hat{\rho}_\tau(x)+\xi_\tau(x)\, ,
$$
with $D=2\sum_{h=1}^M\frac{J_h^2\,h^2}{\gamma}$, and $\xi_\tau(x)$ a noise with properties already discussed. The same result holds if one considers $\varphi \neq 0$, i.e. with $D=\varphi+\frac{2\sum_{h=1}^MJ_h^2\,h^2}{\gamma+2\varphi}$.

\section{Derivation of the structure factor}
\label{AppSF}
In this section we provide details on the computation of the dynamical structure factor. 
As pointed out above, in the large $|\lambda|$ regime one has that the stationary optimal profile $\rho_{opt}(x)$ tends to the value $\rho_{opt}\to1/2$ almost everywhere, except for vanishingly small regions at the boundaries. As a consequence $\sigma(\rho_{opt})\to\sigma=1/2$, so that $\sigma^\prime\to0$. Moreover $\bar{\rho}_{opt}$ is such that $\partial_x\bar{\rho}_{opt}\to0$ \cite{Imparato2009}. Thus, one can approximate $\mathcal{L}_2$ with 
\begin{equation}
\mathcal{L}_2\sim\hat{\mathcal{L}}_2= i\delta\bar{\rho}\left(\partial_{\tau}\delta\rho-D\partial_x^2\delta \rho\right)+
\frac{\sigma}{2}\left(\partial_x\delta\bar{\rho}\right)^2
-\frac{\sigma^{\prime\prime}}{4}\lambda^2(\delta \rho)^2, 
\label{approxsecordbarL}
\end{equation}
with $\mathcal{L}_2=\hat{\mathcal{L}}_2$ only in the $L\to\infty$ limit, for $s\neq0$. Notice that $\hat{\mathcal{L}}_2$ is the exact second order expansion of the Lagrangian $\mathcal{L}$ in the equilibrium case  $\varrho_0=\varrho_1=1/2$. Therefore, for large $L$, the expectation in the s-ensemble is well approximated by
\begin{equation}
\langle O\rangle_s\sim\frac{\int{D}\rho D\bar{\rho}O[\rho]\e^{-L\iint dxd\tau \hat{\mathcal{L}}_2[\delta\rho,\delta\bar{\rho}]} }{\int {D}\rho D\bar{\rho}\,\e^{-L\iint dxd\tau \hat{\mathcal{L}}_2[\delta\rho,\delta\bar{\rho}]}}\, ,
\label{approxflucsensemble}
\end{equation}
which can be computed as we show in the following. Through the space-time Fourier expansion,
$$
\delta \rho_\tau(x)=\frac{\sqrt{2}}{T}\sum_\omega\sum_{p>1}\sin (p\, x)\e^{-i\omega \tau}\delta\tilde{\rho}_{{p},\omega}
$$
with $\omega=\frac{2\pi}{T}u,\, u\in \mathbb{Z}$, and, the equivalent one for $\delta\bar{\rho}_\tau(x)$, one can diagonalise $\hat{\mathcal{L}}_2$, obtaining 
$$
\iint dxd\tau\hat{\mathcal{L}}_2=\frac{1}{T}\sum_{{p}>1,\omega\ge0}\delta\vec{\tilde{\rho}}_{{p},\omega}^{\,\dagger}\cdot K_{{p},\omega}\cdot\delta\vec{\tilde{\rho}}_{{p},\omega}\, ,
$$
where $\delta\vec{\tilde{\rho}}_{{p},\omega}=(\delta\tilde{\rho}_{{p},\omega},\delta\tilde{\bar{\rho}}_{{p},\omega})^{tr}$ (with $a^{tr}$ denoting vector transposition and $a^{\dagger}=(a^*)^{tr}$), and 
$$
K_{{p},\omega}=\begin{pmatrix}
-\frac{\sigma^{\prime\prime}\lambda^2}{2}&\omega+iD{p}^2\\
-\omega+iD{p}^2&\sigma {p}^2
\end{pmatrix}\, .
$$
At this point, the computation of ${\cal S}(p,\tau)$ is reduced to evaluations of Gaussian path integrals. In terms of the Fourier fields $\delta\tilde{\rho}_{{p},\omega}$,  
\beq
\delta\tilde{\rho}_{\tau}(p)=\frac{1}{T}\sum_{\omega}\e^{-i\omega\tau}\delta\tilde{\rho}_{{p},\omega}\, ,
\label{fourtr}
\eeq
one can write
$$
\mathcal{S}(p,\tau)=\frac{1}{T^2}\sum_{\omega,\omega'}\e^{i\omega\tau}\langle \delta\tilde{\rho}_{{p},\omega'}\delta\tilde{\rho}^*_{{p},\omega}\rangle_s\, .
$$
Then, evaluating the above $s$-ensemble expectation with \eqref{approxflucsensemble}, one gets
$$
\langle \delta\tilde{\rho}_{{p},\omega}\delta\tilde{\rho}^*_{{p},\omega^\prime}\rangle_s\sim \delta_{\omega,\omega^{\prime}}\frac{T}{L}\frac{\sigma{p}^2}{\omega^2+D^2{p}^4-\frac{\lambda^2\sigma^{\prime\prime}\sigma{p}^2}{2}}\, .
$$
Thus ${\cal S}(p,\tau)$ reads
$$
\mathcal{S}(p,\tau)\sim\frac{1}{LT}\sum_{\omega}\e^{i\omega\tau}\frac{\sigma{p}^2}{\omega^2+D^2{p}^4-\frac{\lambda^2\sigma^{\prime\prime}{p}^2}{2}} \, .
$$
By replacing the summation over $\omega$ with an integration in the long-time limit, $\mathcal{S}(p)$ reads
$$
\mathcal{S}(p,\tau)=\frac{1}{2\pi L}\int_{-\infty}^\infty d\omega\e^{i\omega\tau}\frac{\sigma{p}^2}{\omega^2+D^2{p}^4-\frac{\lambda^2\sigma^{\prime\prime}{p}^2}{2}}\, .
$$
Therefore, one finds
\begin{equation}
\begin{split}
\mathcal{S}(p,\tau)=&\frac{1}{L}\frac{\sigma{p}^2}{\sqrt{4D^2{p}^4-2\lambda^2\sigma^{\prime\prime}\sigma {p}^2}}\times\\
&\times \exp\left(-\frac{\tau}{2}\sqrt{4D^2{p}^4-2\lambda^2\sigma^{\prime\prime}\sigma p^2}\right)\, .
\end{split}
\end{equation}
Recalling that $\lambda=sL$, $p=Lk$, and that in the large $L$ limit approximations \eqref{approxsecordbarL}-\eqref{approxflucsensemble} become exact, one has 
\begin{equation}
S(k,t) 
= \sigma k^2
\frac{\exp\left(-\frac{t}{2} 
\sqrt{4 D^2 k^4-2 s^2\sigma^{\prime\prime}\sigma k^2}\right)
}
{\sqrt{4 D^2 k^4-2 s^2\sigma^{\prime\prime}\sigma k^2}}\, .
\end{equation}

\bibliographystyle{apsrev4-1}
\bibliography{qhyperuniform}

\end{document}